\documentclass [twoside,fleqn]{article} 
\usepackage{epsfig,espcrc2}

\title{FELIX\\ 
       A full acceptance detector at the LHC}       
 \author{
 V. Avati, K. Eggert,\address{CERN, Geneva, Switzerland}
 and C. Taylor \address{Dept. of Physics, Case Western Reserve University, Cleveland, OH
          44106 USA} \\
 (for the FELIX Collaboration) \\
Presented by K. Eggert}

\begin{document} 
 
\begin{abstract}
The FELIX collaboration has proposed the construction of a full acceptance
detector for the LHC, to be located at Intersection Region 4, and to be commissioned
concurrently with the LHC.  The primary mission of FELIX is QCD:  to provide
comprehensive and definitive observations of a very broad range of strong-interaction
processes.  This paper reviews the detector concept and performance characteristics,
the physics menu, and plans for integration of FELIX into the collider lattice and
physical environment.  The current status of the FELIX Letter of Intent is discussed.
\end{abstract}

\maketitle 

\section{Introduction}

 \begin{figure}[htbp] 
 \begin{center} 
\epsfig{file=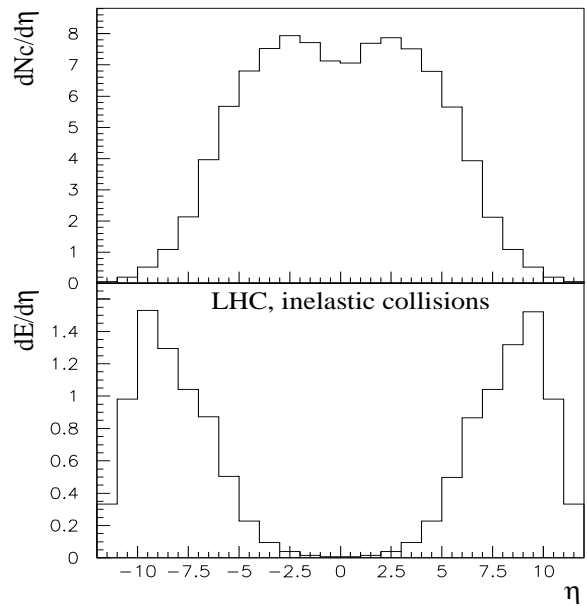,height=8.3cm,width=8cm,bbllx=90,bblly=232,bburx=475,bbury=682,clip=}
\end{center}
\caption{The pseudorapidity distribution of charged particles and of the
energy-flow at $\sqrt{s}=$14 TeV.}
\label{prodplot}
\end{figure}
\begin{figure*}[htb]
\centering\epsfig{file=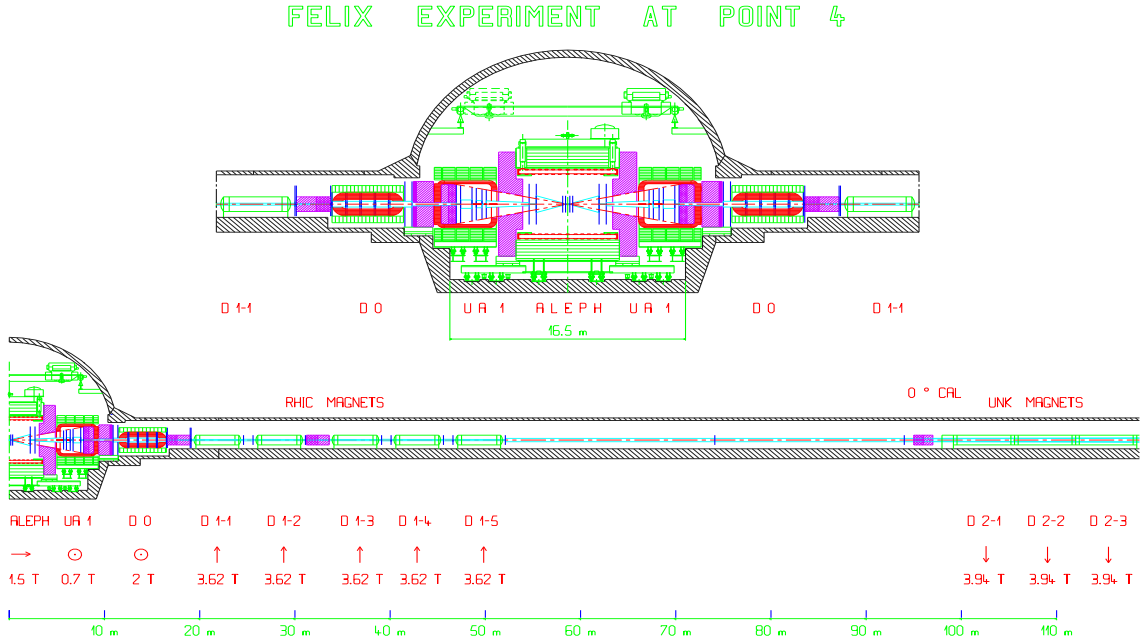,width=10cm,height=16.3cm,angle=-90}
\caption{Sketch of the FELIX experiment in central and very forward region.}
\label{FIGI}
\end{figure*}

FELIX will be the first full acceptance detector
at a hadron collider.  It will
be optimized for studying the structure of 
individual events over all of phase space (see Figure \ref{prodplot}).
FELIX will observe and measure all charged particles, from the central
region all the way out to diffractive protons which have lost only
$0.2\%$ of their initial energy. It will even see elastic
protons which have a momentum transfer of at least $10^{-2}$ GeV$^2$.
This 
comprehensive, precision
tracking is accompanied by equally 
superb
electromagnetic
and hadronic calorimetry.  FELIX will observe and measure photons and neutrons
down, literally, to zero degrees, giving it an unparalleled ability
to track the energy flow.  
In contrast, the other LHC detectors are sensitive over only a fraction
of phase space and see less than 10\% of the typical energy flow.
FELIX is thus uniquely able to pursue physics
complementary to that of the other detectors planned for the LHC.

The FELIX design involves the coordinated arrangement of three distinct
systems:  the magnetic architecture responsible for getting the beams
through the I4 straight section, the  tracking system, and the 
calorimetry.  Each system must be complete in its own right, without
compromising the characteristics of the other systems.  The magnetic
apertures must not be limiting apertures of either the tracking or
calorimeter systems.  
There must be sufficient
physical space for both tracking and calorimetry.  The calorimeters
must be physically large enough to have good resolution, and must not
interfere with either the tracking or the magnetic systems.  

All of this requires
a lot of space, and the detector must be carefully 
integrated into the design of the machine.  
Full acceptance cannot be achieved by ``adding on" to central detectors
optimized for high $p_T$ physics.  Here FELIX is fortunate.
The decision to split the RF cavities at I4, moving them to $\pm$ 140 m
from the interaction point (IP),  
combined with the fact that FELIX's ``low" luminosity permits the
focusing quadrupoles to be moved more than 120 m from the IP, provides
the necessary longitudinal space.
I4 is also ideal from the point of view of transverse space.
The beams are separated by
42 cm at the location of the RF cavities, providing 
room for zero degree calorimetry.  Since the existing infrastructure, 
including the ALEPH solenoid, can be re-used with minimal modifications,
I4 is clearly a superb location for a full acceptance detector.
(The central part of FELIX, which nicely fits into the existing 
cavern, and the extensions upstream into the forward regions,
are shown in Figure~\ref{FIGI}.)

Nevertheless, the task of integrating a detector with 
genuinely full acceptance into the available space at I4 is not trivial.  
The FELIX Letter of Intent \cite{LOI}
outlines how it can be done, using well-understood
magnets and compact detectors, for a comparatively modest price:
we estimate a cost of about 25 MCHF for the 
machine magnets and the infrastructure,
and about 50 MCHF for the detector outlined here and presented in 
more detail in the FELIX LoI.
 
\section{Physics Overview}

The heart of the FELIX physics agenda is QCD: FELIX will be the ultimate
QCD detector at the LHC. 

Surprisingly, the need for such a detector is not obvious to many members of
the high energy community.  In part, this may be because of the success of
the interplay between theory and experiment in the case of electron-positron
collisions.  The
cleanliness of the process, together with the low event
rate and full-acceptance capability
of the detectors, has led to an especially fruitful interaction between
the QCD aspects of that experimental program with the remainder. 

The case of hadron-hadron collider physics is quite different. The
high-$p_T$, low cross section physics is accessed by highly selective
triggers. The phase-space acceptance of the detectors is largely limited to the
central rapidity region. Full acceptance has not been attained since the
bubble-chamber era of fixed-target physics. Therefore the basic data base
is much more limited. 

This situation is all the more serious because of the great
variety in event classes for hadron-hadron collisions. There are soft
collisions with large impact parameters;  angular momenta of tens of
thousands instead of the unique $J = 1$ of the $e^+e^-$ world. Central
collisions produce much higher multiplicities than  are seen in
$e^+e^-$ annihilation. There are the diffraction classes of events, with
and without jet activity, that comprise several to tens of percent of
typical subsamples (if seen in full acceptance) and which present a major
challenge to theory. There are poorly understood strong Bose-Einstein-like
correlations seen at very low $p_T$ and low relative $p_T$ in
hadron-hadron collisions which do not occur in $e^+e^-$ collisions. But
at
collider energies this is only based on one sample of low-$p_T$ data from
UA1, because until now no other detector has had the measurement
capability. Finally,
there is little if any data in the forward fragmentation regions,
where cosmic ray experiments insistently claim that anomalies exist.

Given this richness of phenomena, and given the importance of QCD to the
interpretation of the new-physics data 
expected to emerge from the LHC, it is
clearly very important to improve the data-base with an LHC detector and
experimental group fully dedicated to the observation and interpretation
of as broad a range of QCD phenomena as possible. This is of course the
mission of the FELIX initiative. 

Many of these new opportunities in QCD physics at
the LHC are not well known, and the FELIX collaboration  has 
accordingly placed high
priority in in providing a description of them in the FELIX LoI.  We 
briefly summarize a few of the main themes here.

\subsection{Parton densities can be measured to extremely small $x$,
below $10^{-6}$}

The parton densities at small $x$ are themselves a very
important thing to measure. Up to now HERA has provided data down to $x$
values of order $10^{-4}$ for $Q^2$ in the perturbative domain of several
GeV$^2$. FELIX will have the 
capability 
to extend
these
measurements to $x$ values below $10^{-6}$ via observation of dileptons,
low-mass dijets, and low-mass jet-photon systems carrying large
longitudinal momenta. In this regime one expects (especially for
proton-ion collisions) the breakdown of the usual DGLAP/BFKL
evolution-equation formalism and significant nonlinear effects to be
observed. 
 
\subsection{Minijet production in hadron-hadron collisions is strongly
energy dependent}

The need for a vastly improved QCD data-base for hadron-hadron collisions 
is made even more urgent by the fact that 
qualitative changes are expected
even in the structure of generic events because of
the rapid increase with energy of gluon parton densities in the primary
protons. Thanks to the measurements at HERA, this is not only the
theoretical expectation but also a data-driven one. The parton densities
at a $5-10$ GeV scale become so large that minijet production in central
collisions may become common place, with minijet $p_T$ large enough for
reasonably clean observability. These very high parton densities create,
at a perturbative short distance scale, ``hot spots'' in the spacetime
evolution of the collision process within which there may be
thermalization or other nonperturbative phenomena not easy to anticipate
in advance of the data.  Particle spectra themselves may evolve to
something quite distinct from what has been so far observed, with
strangeness, heavy flavors, and/or baryon and antibaryon production
enhanced. Especially in central proton-ion collisions, where the total
gluon-gluon luminosity per collision is maximized, and where the evolution
of a single proton fragment is followed, one can expect this class of
phenomena to be most prominent and surprises most probable.

\subsection{Diffractive final states are endemic, many are important,
and some are spectacular}
\label{ov.5}

Diffractive final states
will comprise almost $50\%$ of all
final states at the LHC. 
The soft diffraction at very large impact parameter, which perhaps
sheds light on pion-cloud or glueball physics, is at one extreme, and hard
diffraction, where rapidity gaps coexist with jets, is at the other. 
There are a large variety of hard diffraction processes,
including some with two and three rapidity gaps, which are of basic
interest to study. In this class there are expected to be, for example, an
extraordinary class of events where the complete event consists of a
coplanar dijet accompanied by the two unfragmented beam protons detected
in Roman pots, and absolutely nothing else in the detector. Certainly
ATLAS and CMS can
also detect such events, provided they sacrifice a
luminosity factor of about 30 relative to their hard-earned peak luminosity.
However, to really understand this event class, one will need, at the very
least, to examine the $t$-distribution of the Roman-pot protons, as well
as
to study the generalizations of this process to the cases where one or
both of the protons undergoes soft diffraction dissociation to a low mass
resonance or a high mass continuum, or to a high-$p_T$ system containing a
tagging jet. Only FELIX would have such a capability.

In addition to this class of hard diffraction and very soft diffraction
processes, there is another very interesting class of semihard diffractive
phenomena associated with the conjectured fluctuation of the initial-state
projectile into a transversely compact configuration, which therefore
interacts with an unusually small cross section. Evidence for this is seen
in vector-meson photoproduction at HERA, especially $J/\psi$ production,
which
exhibits the expected rapid increase of cross section with energy. Also at
Fermilab, diffraction dissociation of a high energy pion into dijets, with
all the initial pion energy going into the dijet system, is being studied by
experiment E791. Exactly the same process is available at the LHC
with FELIX, as well as a similar process where one beam proton dissociates
diffractively into three jets, one for each quark. The $A$ dependence of
these processes is remarkable, roughly $A^{4/3}$, because this diffractive
process should occur even in central collisions, thanks to the small size
of the initial configuration.

 
\subsection{Particle production from deep within the light cone may
exist and deserves careful searches}

The existence of events with a very high final-state multiplicity of
minijets and their associated hadrons has other implications. The products
of such interactions for the most part can be expected to explode from the
initially compact collision volume in all directions at the speed of
light. Because of the high multiplicity density, the time of hadronization
of all these degrees of freedom will be lengthened from the usual
low-energy value of 1-2 fm to several fm. Up to this time of
hadronization, the expanding ``fireball'' containing most of the partonic
collision products is arguably a rather thin spherical shell, of thickness
of order a fm. So even before hadronization there is a large interior
volume of hundreds of fm$^3$, isolated from the exterior
vacuum, which may evolve toward a chirally disordered vacuum.
Consequently in such events there might be a large pulse of semiclassical,
coherent pions of relatively low $p_T$ emitted when this false vacuum
eventually decays: disoriented chiral condensate. This is at present only
a speculative possibility, although experimental searches, especially in
the context of ion-ion collisions, are underway. 

More generally, one may
ask: if disoriented vacuum is not what is in the interior of this
quasi-macroscopic fireball, what is? If the interior ``vacuum" is broken
into domains of various chiral orientations, then topological obstructions
might
lead to production of (Skyrmionic) baryons and antibaryons of unusually
low $p_T$. And if there is activity
deep inside the light
cone, no matter what it is, 
then this activity has  eventually to be turned into emission of
particles; hence a new particle production mechanism which deserves to be
studied. It would seem that the only alternative available for the
$absence$ of new phenomena emergent from the deep interior of the light
cone under these circumstances is that that region relaxes back to the
true vacuum, despite its being isolated from the true vacuum by a
fireball shell and despite there not being  enough elapsed time for
chiral orientation to be distinguished energetically 
from chiral disorientation.

\subsection{Collisions with very high impact-parameter may probe
the chiral
vacuum structure}

In general, the chiral vacuum condensate is distorted in the neighborhood
of impurities such as an isolated proton. This is just the long-range pion
cloud surrounding it. The pion-cloud structure can be probed especially
well in high energy $pp$ collisions at very large impact parameters,
say 2 to 3 fm. These interactions are, because of the larger radii of
interaction at the LHC, a bigger component of the cross section, and can
lead to larger final-state multiplicities than found at lower energies. 
Perhaps here too there may be coherence in the structure of the pion
emission, and this class of events may turn out to be of special interest.
Again a detection capability at very low $p_T$, 100 MeV and less, as
possessed by FELIX, is important for such studies.

\subsection{New opportunities exist for tagging event classes} 
\label{ov.8}

Together with these many novel phenomena, there will be new methods for
experimentally tagging different kinds of events. The impact parameter of
the collision is obviously of importance to be determined event-by-event. This
is done routinely in ion-ion collisions via zero degree measurements of
nuclear fragments and by the amount of transverse energy produced. At the
LHC, the FELIX instrumentation in the forward direction allows a
data-driven approach for attacking the problem by the former method. The
large yield of minijets, strongly dependent upon impact parameter, may
allow the latter method, based upon transverse energy production, to be
used more effectively at the LHC (by all detector groups) because of the
stronger correlation of multiplicity with impact parameter than at lower
energy.  A combination of both methods, unique to FELIX, is likely to be
the best of all.
 
A second important tag available to FELIX is the choice of beam. By
tagging on a leading neutron or $\Delta^{++}$ at very low $t$, one can
reasonably cleanly isolate the one-pion-exchange contribution, and
thereby
replace the LHC $pp$ collider with a somewhat lower energy, lower
luminosity $\pi p$ collider.  In a similar spirit, and including $\Lambda$
tags, one can study collisions of any combination of $\pi$, $K$, or $p$
with each other. The beam-dependence of phenomena has historically been of
considerable importance, and it  may
find important applicability, especially with respect to questions of
valence-parton structure, at the LHC energy scale.

A special case of these tags is that of a photon tag in ion-ion
collisions,
via forward detection of the undissociated ions. 
The luminosity for $\gamma \gamma$ collisions is very high, and the
capability of FELIX to exploit this luminosity is also very high. 

Another class of tags which has been underutilized is the diffractive
tag, where leading protons are detected via Roman-pots. As discussed
above, this leads to a very rich stratum of up-to-now
poorly-measured, poorly
understood, but potentially important physics.

  \begin{figure*}[htb]
\begin{minipage}{\textwidth}
 \epsfig{file=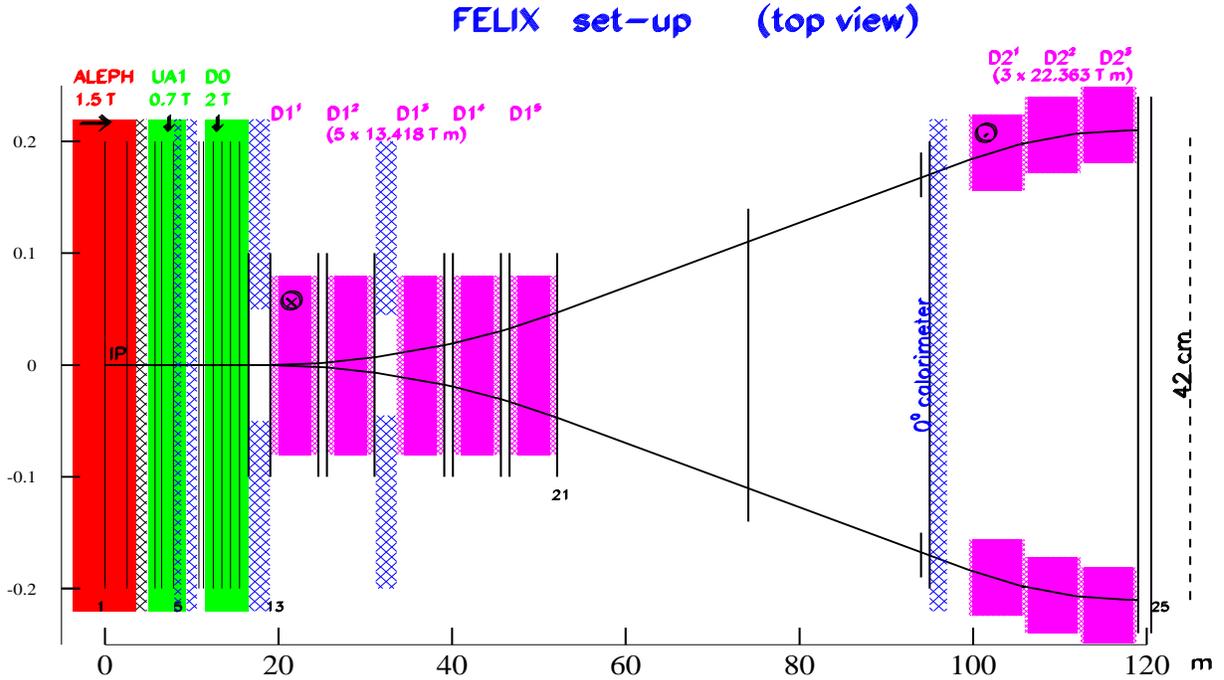,bbllx=57pt,bblly=54pt,bburx=516pt,bbury=790pt,%
 width=9cm,height=16cm,angle=-90,clip=}  
\caption{The top view of the FELIX detector.  The  different
magnets, calorimeters (hatched areas), tracking stations (vertical lines)
 and the beam trajectories
in the horizontal plane are indicated.}
\label{setupvfnewi}
\end{minipage}
\end{figure*}
 
Finally, there may be pattern tags. The event structure in final states
containing jets is dependent upon the color flow. Typically, neighboring
jets in phase space are connected by a partonic color line (antenna). For
quarks, one antenna line emerges from the jet, for gluons two. Along these
antenna lines in phase space, hadronization and minijet production is
enhanced. Recently the Tevatron collider experiments have observed these
effects.  In principle this technique might allow one in the future to
identify in an individual multijet event quarks versus gluons, and even
fully classify the event structure according to the color flow.  Clearly
such a pattern-analysis technique is very difficult, and needs to be
data-driven. FELIX, with full acceptance, will be
optimal for making the attempt. 
  
\section{The FELIX detector}

We  now introduce
the major features of the FELIX design. More details can be found in the FELIX
LOI.~\cite{LOI}.

\subsection{A tunable insertion at I4}

A full acceptance detector must be able to analyze the global
structure event-by-event.  This means that it should run at a luminosity
no greater than ${\cal L} {\sim} 10^{32}$ cm$^{-2}$ s$^{-1}$; that is, with
less than about one interaction per crossing. 
This luminosity can be achieved at I4 by
means of an insertion which can be tuned from $\beta^* = 23$ m to
$\beta^*=900$ m without changing the magnetic elements.  
 
There are two significant features of this insertion.  First, 
the final-focus quadrupoles can be placed more than 120 m from
the IP, providing the space needed to accommodate the FELIX
dipoles.  Second, it is economical.  The necessary quadrupoles are
already in the LHC baseline design.

The ability to tune the insertion also has several nice features.
At $\beta^*=900$ m, FELIX is optimized for the study
of low-t elastic scattering. 
At $\beta^*=110$ m, where FELIX's luminosity is about 
4 x 10$^{31}$ cm$^{-2}$ s$^{-1}$ when
the LHC is at design luminosity, the beam
size in the heart of FELIX detector ($\pm$ 120 m) is minimized, permitting
the Roman pot detectors in these locations to come as close
as 3 mm to the beam.  
Finally, $\beta^*=$ 23 m permits FELIX to reach luminosities
as high as 2 x 10$^{32}$cm$^{-2}s^{-1}$.

\subsection{Well-understood magnets}

FELIX will implement a ``kissing scheme" 
in which the two beams are brought together 
at $0^o$ in the horizontal plane and
then returned to the same inner or outer arc (See Figure \ref{setupvfnewi}). 
To accomplish this, we
need  
some 67 T-m 
to first bring the beams together
(D2 magnets), and then another 67 T-m (D1 magnets)
to make them parallel.  This has to be accomplished  within the
120 m available. Both sets of magnets must be superconducting
machine dipoles.  The D1 magnets must also have large bores, to
accomodate both  beams and  to provide acceptable tracking and calorimetry
apertures.

FELIX is fortunate that Brookhaven National Laboratory (BNL) has
designed  large aperture superconducting dipole magnets for use at RHIC.
With a coil aperture of 18 cm and a design field
of 4.28 T (FELIX will use them at 3.62 T), these magnets are suitable
for use as D1 magnets.  
BNL is committed
to producing these magnets for RHIC and thus will be able to supply 
well-understood magnets on the FELIX time scale.

The constraints on the D2 magnets are somewhat less severe, and several
options are available.  Of these, FELIX proposes existing superconducting
dipoles constructed as prototypes for UNK.  While these are
single aperture magnets, the 42 cm beam separation  permits two UNK cold
masses to be assembled in a common cryostat for use as D2 magnets.

In order to avoid parasitic beam-beam interactions and long-range
tune shift effects,  the beams will collide with a vertical
crossing angle of $\pm 0.5$ mrad.  To do this while optimizing the match
of the magnetic architecture to tracking and calorimetry, we propose to
re-use the existing UA1 magnet, split longitudinally into two halves
and equipped
with new coils. We will also  build two 
5 meter long, 
2 T warm dipole (D0) magnets.

The  magnetic architecture is completed by the re-use of the
existing ALEPH solenoid, which is well-matched with the use of the UA1
magnet.

An important feature of this overall design is that the strengths of
the magnetic fields increase in the  forward direction, always well-matched
to the typical momenta of the particles, resulting in momentum resolution
which is reasonably uniform over all of phase space.

Finally, we note that all magnets can be accommodated in the existing
Aleph collision hall and adjacent tunnels without any significant civil
construction.

\subsection{Compact, precise tracking}

\begin{figure}[htbp]
\centering
 \epsfig{file=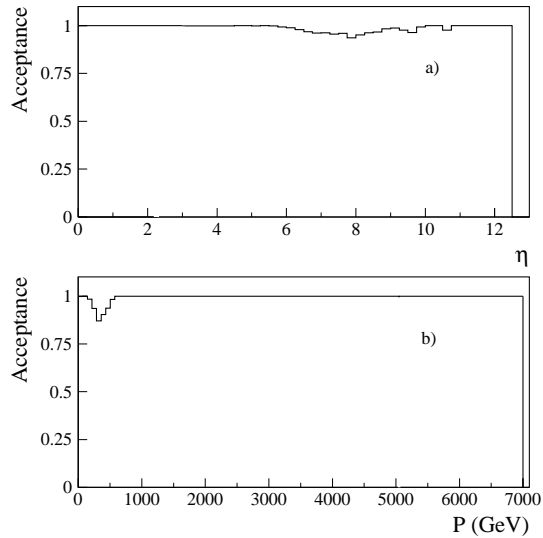,width=8cm,height=8cm} 
\vspace{-0.5cm}
\caption{The acceptance in FELIX for charged particle momentum measurements 
as a function of (a) the pseudorapidity; and (b) the momentum of the particles.} 
\label{acc-fract}
\end{figure}

Some 50 tracking stations, located as far as 430 m from the IP, are
needed to ensure full acceptance and uniform resolution. 
The positions of most of the stations (vertical lines) are indicated in Figure
\ref{setupvfnewi}.
FELIX will instrument radially outward,
emphasizing compact, near-beam tracking. How close we will approach
the beams depends on the location.
In general, we will use Roman pot detectors to aggressively approach the beams wherever the location
is accessible and the pot mechanical structure does not interfere
with other tracking or calorimetry.  Elsewhere, we propose to use 
fixed-radius tracking, approaching to within 2.5 cm of the beams. 
%
The acceptance for charged particles as a function of pseudorapidity
(a) and their momenta (b) (see Figure ~\ref{acc-fract}) is almost 100\%
over the entire phase space.

An important consideration is the occupancy within the tracking detectors.
High particle densities close to the beam
 pose a significant pattern recognition
problem.  Each tracking station should thus have sufficient resolution
and redundancy to be able to locally reconstruct track segments.  Track
segments are then matched, station-to-station, resulting in a very
powerful spectrometer.

\begin{figure}[htbp]
\epsfig{file=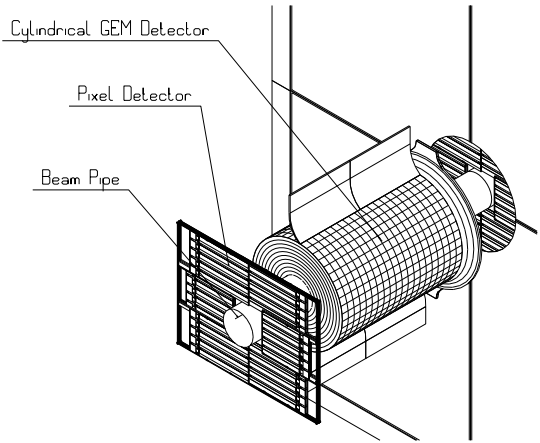,
bbllx=10pt,%
bblly=40pt,bburx=580pt,bbury=800pt,
width=6cm,angle=-90
} 
\caption{A schematic view of a tracking station based on Si pixel
detectors and a micro-TPC.  Note that several 
large-area GEM chambers have
been removed to improve visibility of the micro-TPC.}
\label{BEAM_CYL_009i}
\end{figure}

These considerations lead to a common conceptual design for most FELIX
tracking stations, based on two technologies:  Si pixel detectors 
out to radii of about 8 cm, supplemented by Gas Electron Multiplier
(GEM) chambers at larger radii.  We are also exploring the possibility
of using GEM as the basis for very compact micro-TPC's.  
A conceptual design for a ``standard" fixed-radius tracking station
is shown in Figure  \ref{BEAM_CYL_009i}.
The same
technologies will be used for a compact microvertex detector.

\subsection{Forward calorimetry}
\begin{figure*}[htb]
\epsfig{file=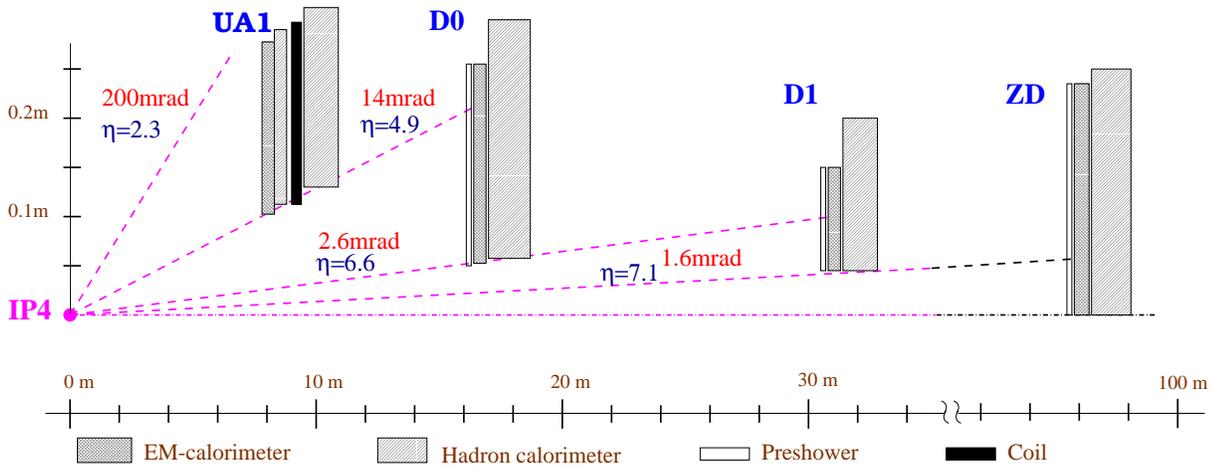,width=16cm}
\caption{Schematic view of FELIX forward calorimetry.}
\label{FELIXcalsi} 
\end{figure*}

FELIX proposes four calorimeters on each side of the IP to provide
complete electromagnetic and hadronic calorimetry for angles 
$\theta < 0.2$ radian, that is, for $|\eta|>2.3$.  The coverage
of the calorimeters is illustrated
in Figure \ref{FELIXcalsi}. The interplay with the magnets and tracking
systems is illustrated in Figures \ref{FIGI}.

The calorimeters must have superb energy and spatial resolution,
and must provide the information needed to identify neutrons, electrons
and gammas.  This must be done in limited space, and in a high-radiation
environment. These considerations  determine the structure of the
calorimeters, the choice of sampling materials and the kinds of
photodetectors and front end electronics which can be used for the readout.

The UA1 endwall calorimeter, which is expected to have a radiation dose
of less than  5 Mrad for 10 years running, is a sampling calorimeter
based on plastic scintillators and wavelength shifting fibers. 
The
very forward (D0, D1 and Zero Degree calorimeters) see much higher radiation
levels, and will thus be ``spaghetti"-type calorimeters, based on either
thin capillaries filled with liquid scintillator or on quartz fibers.
All three very forward calorimeters
are similar in construction, differing only in their overall
dimensions.  Each consists of a preshower detector, an EM calorimeter,
and two hadron calorimeter sections.  

\subsection{Trigger }

\begin{figure}[htb]
\epsfig{file=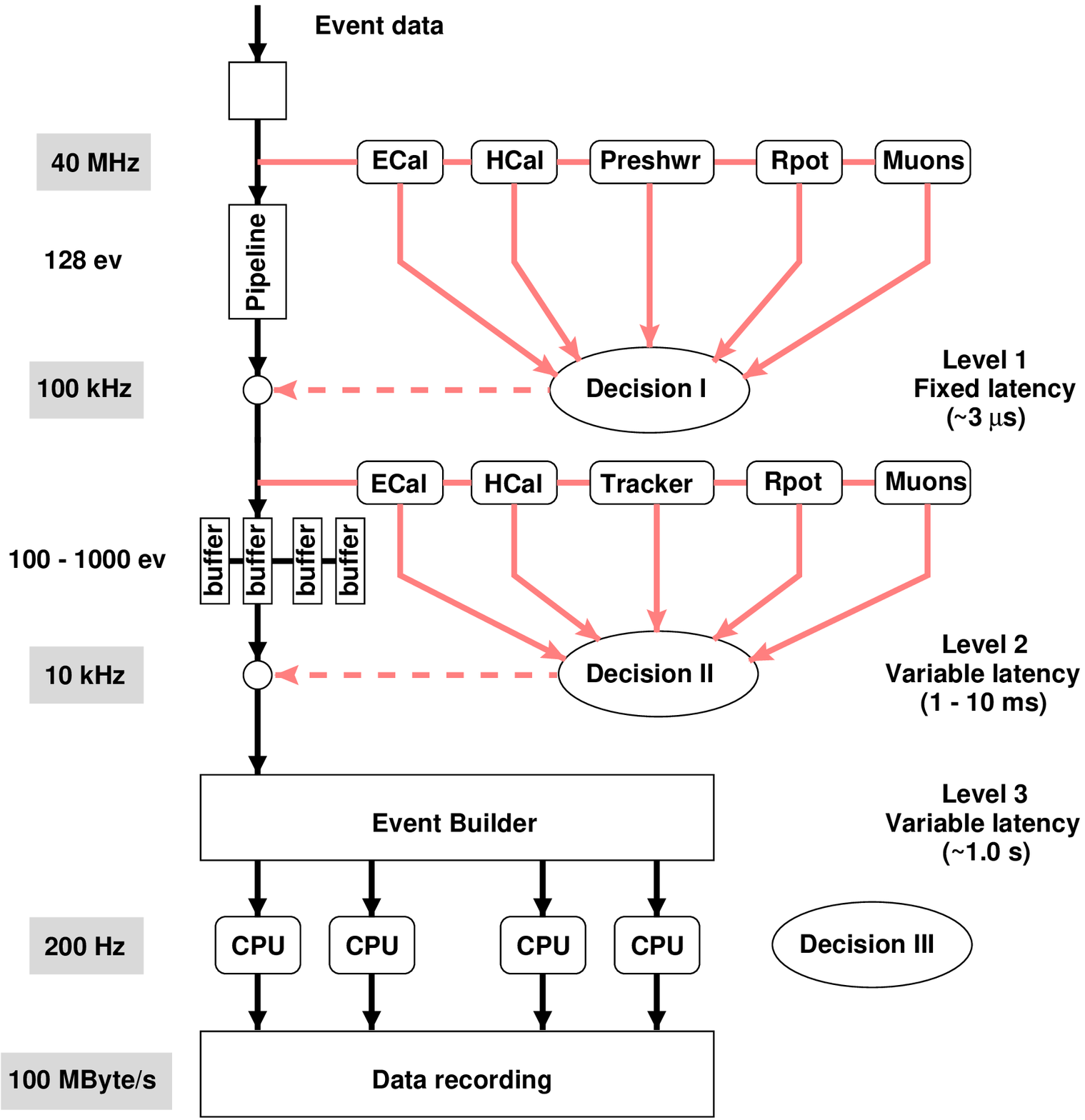,width=8cm,height=8cm}
\caption{The FELIX triggering and DAQ scheme.}
\label{trig} 
\end{figure}

The basic structure of the trigger is a multi-level triggering scheme which must reduce
the trigger rate from 40 MHz to a rate acceptable for data recording. A schematic view of
the trigger scheme together with the rates and the latencies of the three trigger
levels is shown in Figure~\ref{trig}.

The initial Level 1 triggering must be done deadtimeless for each bunch
crossing.  It must establish clean evidence of the presence of an
event from the triggering constituents and produce a
constituent-based trigger decision.  This is achieved with simple
algorithms based upon the topology of the elements of the detectors
which fired, or upon special trigger counters and energy deposited
therein.  These local trigger decisions are produced near the
detector front-end electronics by special circuitry.
During the
Level 1 latency, the sampled data of all channels of all detectors
are stored in data pipelines.
After a positive Level 1 trigger decision, the output data are
associated with
specific bunch crossings, formatted into event fragments, and
stored for subsequent filtering by the Level 2 trigger.
The output
rate from Level I is estimated to be around 100 kHz. 
 
The Level 2 system performs a fast decision on the event quality,
based upon data from separate detectors or portions thereof, using
more sophisticated algorithms. These may be realized by a set of
dedicated hardware or software 
processors. The output rate from Level 2 is estimated to be roughly
10 KHz.

The Level 2 filtered data are combined into complete events by the
event builder, and are filtered by the Level 3 computer
farm.  Level 3 produces the complete event-based
decision, and its task is to reduce the trigger rate from 10 KHz to
one acceptable for data recording (around 100 MBytes/sec). We
estimate the mean FELIX event size to be around 0.5 MBytes. 
Therefore the Level 3 rate might be around 200 Hz.

\subsubsection{Run scenarios}

The first major data set containing inelastic events used for
serious physics analyses is assumed to be of order $10^8$ $pp$
collisions, taken with essentially a minimum-bias trigger. The only
selectivity provided at trigger level would address the rejection
of bunch crossings with no interactions (or more than one
interaction), and rejection of candidate events seriously
contaminated with beam-gas background. 

After such a minimum-bias run, an intermediate-luminosity run could
be naturally implemented using only the Level 2 trigger strategy
and a choice of interaction rate compatible with the lack of
rejection that Level 1 would provide. Such a strategy would seem to
maximize the flexibility of choice of detailed triggers, and leave
to a minimum the amount which is hard-wired into the front end of
the DAQ system. We assume here that this intermediate-luminosity
run might yield of order $10^8$ recorded events for $10^{11}$ $pp$
interactions in the FELIX detector, i.e.  have a rejection of order
$10^3$. 

Finally, the design run, fully implementing the Level 1 and Level
2 triggers, should be expected to yield $10^8$ recorded events per
$10^{14}$ interactions in the detector.

The detailed choice of the trigger algorithms should probably be
deferred for as long as possible, for at least two reasons. One is
technological; the rate of change is so large that one should opt
for as modern hardware (and software) as possible. The other is the
physics itself, which is evolving and will continue to evolve
rather rapidly with time. It is not easy to anticipate in detail
what will be the highest-priority QCD physics to address at the
time of LHC commissioning.


\section{Recent History}

After the presentations of J.D. Bjorken and K. Eggert about possible forward 
physics in pp and p-A collisions at the LHCC "Workshop on Further Physics Topics" (Nov. 1994)
the LHCC Committee recommended this kind of physics by noting:
{\it
   ``The LHCC noted the interest in diffraction, 
	    and expects that such studies may also form 
	    part of the LHC experimental programme. 
	   The committee encourages interested parties   
	    to work together on an integrated approach  
	    towards this physics, whilst bearing in mind 
	    the LHC physics priorities already established''} 
 
When the possibility for a new interaction region in I4 became reality (summer 1995)
several workshops took place to discuss the layout of a full acceptance detector.

In May 1996 the LHCC defined new rules for coming activities :
{\it
`` The LHCC urges that any new experimental initiative should be consistent with the
restricted resources likely to be available, and combined as far as possible with one
of the foreseen experiment."}

In an Oct. 1996 memorandum\cite{memo} to the LHCC, FELIX responded to these new guidelines by 
describing, in detail, the FELIX set-up, strategy and financial assumptions. 
The group received general
encouragement from the CERN management to go ahead with the Letter of Intent.

During the spring and summer of 1997, the FELIX collaboration  mobilized
for the preparation of the LoI, which was submitted to the LHCC
in August 1997.

In November 1997, the LHCC chose to address the FELIX 
Letter of Intent, finding

{\it
... that the FELIX LoI is not
responsive to these guidelines. While the physics topics addressed by the programme proposed
in the LoI are of interest (particularly the complete reconstruction of diffractive events), the likely
costs of constructing the proposed dedicated detector and of the modifications to the LHC
collider are very high in comparison with the probable physics output. Finally, the composition
and strength of the collaboration seem inadequate for carrying out a strong programme
addressing these physics topics. \cite{LHCC}}

The CERN Research Board has since endorsed the decision of the LHCC.

The FELIX collaboration believes
that these decisions were reached in a precipitate manner, 
with gross violation of due
process. In particular, there has been no thorough scientific review of
the FELIX proposal. Indeed, a
primary grievance is that the LHCC referees never contacted the proponents
before arriving at its negative conclusion, nor were the proponents
permitted to directly present the initiative in person to the committee.
Important issues, including possible staging scenarios to reduce cost, and
ongoing efforts to build collaboration strength, were thus never presented
to the committees. 

The justification of the decision
 which has been presented by the LHCC, the Research Board and by the Director
General clearly has to do with costs: CERN is under great financial
stress, and the issue of affordability is of course a very real one, an
issue not unnoticed by the collaboration. It is clear that the FELIX
collaboration as presently constituted is far from being able to
provide the resources, a point which was reinforced in private discussions by the
CERN Director General, who has indicated that he might 
have considered the FELIX LoI more seriously if the Collaboration would have 
been stronger, and with more collaborators from CERN Member States.

FELIX has formally protested both the conclusions of the LHCC and
the procedure by which the FELIX LoI has been considered
by the LHCC.

The LHCC and Research Board have, 
however, raised several critical points. FELIX had originally expected to
address such issues via direct interaction with the referees and the LHCC through
the usual procedures. In the present situation, we believe that the best
way of proceeding is to present an addendum to the FELIX LoI to the LHCC
which will contain a thorough discussion of the following points:

\begin{itemize}
\item the complementarity of the capabilities of FELIX with those 
of the already
foreseen experiments;
\item staging scenarios for the FELIX detector; including
\item the possibility to construct a preliminary version of the FELIX 
   experiment at FNAL, HERA or RHIC with a stronger collaboration to demonstrate 
   both the technical feasibility as well as to obtain a first glimpse of the
   physics.
\end{itemize}

FELIX welcomes all additional collaborators, and
will continue to expand the collaboration,
with particular emphasis on CERN member states.  FELIX will also work
to identify funding sources.  Finally, the entire FELIX 
collaboration will continue to work on
substantive issues as outlined above, and in the LOI.  
In particular, we are proceeding with the design and construction of
a prototype forward tracking station, 
as sketched in Figure~\ref{BEAM_CYL_009i},
to be tested at one of the current generation of colliders. 
Prototypes of the various forward calorimeters are
also under construction.

FELIX looks forward to a 
more positive response from the Committees. It is clear, however, that more 
people must soon join the effort if FELIX is to succeed.

\section{Acknowledgements}

Many people have made invaluable contributions to FELIX without formally joining
the collaboration.    We would particularly like to acknowledge the contributions
of the approximately 100 people in this category who directly contributed to the
preparation of the FELIX Letter of Intent, and who are thanked in that document
by name.

\end{document}